\documentclass[aps,twocolumn,prb,longbibliography,showpacs,floatfix,superscriptaddress]{revtex4-1}

\usepackage{graphicx,color}
\usepackage{amsfonts}
\usepackage[figuresright]{rotating}  
\usepackage{amssymb}
\usepackage{bm}
\usepackage{bbm}
\usepackage{physics}
\usepackage{amsmath}
\usepackage{mathtools}
\usepackage{psfrag}
\usepackage{floatrow}
\usepackage{multirow}
\usepackage{tabularx}
\usepackage{textcomp}
\usepackage{units}
\usepackage{lipsum} 
\usepackage{soul}
\usepackage{titlesec}
\usepackage{times}
\usepackage{hyperref}
\usepackage{enumitem}
\usepackage{caption}
\DeclareMathAlphabet\mathbfcal{OMS}{cmsy}{b}{n}

\titlespacing*{\section}
{0pt}{1ex plus .25ex}{1ex plus 1ex}
\titlespacing*{\subsection}
{0pt}{1ex plus .25ex}{1ex plus 1ex}
\titlespacing*{\subsubsection}
{0pt}{1ex plus .25ex}{1ex plus 1ex}

\titleformat{\section}
 {\fontsize{10}{15}\bfseries }
  {\thesection}
  {1em}
  {}

\titleformat{\subsection}
  {\bfseries }
  {\thesubsection}
  {2em}
  {}

\titleformat{\subsubsection}
  {\itshape}
  {\thesubsubsection}
  {2em}
  {}

\def\beq{\begin{eqnarray}}
\def\eeq{\end{eqnarray}}

\let\baraccent=\= 
\renewcommand{\=}[1]{\stackrel{#1}{=}} 
\newcommand{\bk}{\vb{k}} 
\newcommand{\br}{\vb{r}} 

\titleclass{\subsubsubsection}{straight}[\subsection]

\newcounter{subsubsubsection}[subsubsection]
\renewcommand\thesubsubsubsection{\thesubsubsection.\arabic{subsubsubsection}}

\titleformat{\subsubsubsection}
   {\normalfont\normalsize\itshape \centering}{\thesubsubsubsection}{1em}{}
\titlespacing*{\subsubsubsection}
{0pt}{1ex plus .3ex}{1ex plus .3ex}

\renewcommand\paragraph{\@startsection{paragraph}{5}{\z@}%
  {3.25ex \@plus1ex \@minus.2ex}%
  {-1em}%
  {\normalfont\normalsize}}
\renewcommand\subparagraph{\@startsection{subparagraph}{6}{\parindent}%
  {3.25ex \@plus1ex \@minus .2ex}%
  {-1em}%
  {\normalfont\normalsize}}
\def\toclevel@subsubsubsection{4}
\def\toclevel@paragraph{5}
\def\toclevel@paragraph{6}
\def\l@subsubsubsection{\@dottedtocline{4}{7em}{4em}}
\def\l@paragraph{\@dottedtocline{5}{10em}{5em}}
\def\l@subparagraph{\@dottedtocline{6}{14em}{6em}}

\begin{document}
\title{Robust quantisation of circular photogalvanic effect in multiplicative topological semimetals}
\author{Adipta Pal}
\affiliation{Max Planck Institute for Chemical Physics of Solids, Nöthnitzer Strasse 40, 01187 Dresden, Germany}
\affiliation{Max Planck Institute for the Physics of Complex Systems, Nöthnitzer Strasse 38, 01187 Dresden, Germany}

\author{D{\'a}niel Varjas}
\affiliation{Max Planck Institute for Chemical Physics of Solids, Nöthnitzer Strasse 40, 01187 Dresden, Germany}
\affiliation{Max Planck Institute for the Physics of Complex Systems, Nöthnitzer Strasse 38, 01187 Dresden, Germany}
\affiliation{IFW Dresden and W\"urzburg-Dresden Cluster of Excellence ct.qmat, Helmholtzstr. 20, 01069 Dresden, Germany}

\author{Ashley M. Cook}
\affiliation{Max Planck Institute for Chemical Physics of Solids, Nöthnitzer Strasse 40, 01187 Dresden, Germany}
\affiliation{Max Planck Institute for the Physics of Complex Systems, Nöthnitzer Strasse 38, 01187 Dresden, Germany}

\begin{abstract}
Nonlinear response signatures are increasingly recognized as useful probes of condensed matter systems, in particular for characterisation of topologically non-trivial states. The circular photogalvanic effect (CPGE) is particularly useful in study of topological semimetals, as the CPGE tensor quantises for well-isolated topological degeneracies in strictly linearly-dispersing band structures.
Here, we study multiplicative Weyl semimetal band-structures, and find that the multiplicative structure robustly protects the quantization of the CPGE even in the case of non-linear dispersion. Computing phase diagrams as a function of Weyl node tilting, we find a variety of quantised values for the CPGE tensor, revealing that the CPGE is also a useful tool in detecting and characterising parent topology of multiplicative topological states.
 \end{abstract}
\maketitle

 Discovery of the quantum Hall effect~\cite{klitzing1980, PhysRevLett.48.1559} (QHE) has, over the past four decades, expanded into a large branch of condensed matter physics and revealed the central role of topology in this field~\cite{PhysRevLett.50.1395, FQH-Laughlin, kallin1984excitations, PhysRevLett.62.82, halperin1993theory, halperin1982quantized, halperin1984statistics, wen1991gapless, PhysRevLett.63.199, BERNEVIG2002185, nayak2008, sarma_majorana_2015, freedman_two-eigenvalue_2002, Kitaev_2001}. Experimental signatures of the QHE are undeniably topological: a two-dimensional electron gas subjected to an out-of-plane applied magnetic field realizes Hall conductivity quantised to rational numbers $\nu$ in units of $\frac{e^2}{h}$, to one part in a billion. Many symmetry-protected topological phases (SPTs) descending from the integer QHE lack such distinctive signatures~\cite{ryu2010, schnyder2008, PhysRevB.78.195424, Kitaev_2009, altland1997, chiu2016, 3DTI-exp1, chern-discovery, Ersatz-fermi-SPT}, however, and are instead characterized by subtler response signatures or evidence of bulk-boundary correspondence. It is therefore essential to seek out quantised response signatures and better understand the origins of this distinction between the foundational QHE and the large body of symmetry-protected topological phases now studied. 


One of the more recent examples of quantised response in topological phases of matter is the circular photogalvanic effect (CPGE),  a second-order response to circularly polarized light in topological semimetals~\cite{deJuan2017}. For Weyl semimetals~\cite{yanreview2017, jia_weyl_2016, RevModPhys.90.015001, burkov2011, halasz2012time, haldane2014attachment, mathai2017global, hosur2012friedel, potter2014quantum, NIELSEN1983389, zyuzin2012topological, PhysRevB.88.104412, PhysRevX.4.031035, PhysRevX.5.031023, zhang_signatures_2016} (WSMs), the CPGE response is quantised in the limit of small frequency compared to the energy splitting to other bands at the Weyl-node~\cite{deJuan2017}, as additional bands yield unquantised corrections. The CPGE response was also studied for multi-Weyl and multifold fermions~\cite{Flicker2018, Le2020, Le2021, Ni2021}, but quantisation relies on the linearity of the Hamiltonian near the degeneracy point.


To probe the limits of quantised response in topological phases of matter, in this work we therefore consider response signatures of the recently-introduced multiplicative topological phases of matter~\cite{Cook2022}, distinguished by Bloch Hamiltonians with symmetry-protected tensor product structure. We investigate the CPGE in multiplicative Weyl semimetals (MWSMs) specifically as the canonical multiplicative topological semimetals~\cite{Pal2023}. Considering a variety of toy models for MWSMs, we find rich phase diagrams displaying a wide range of quantised values of the CPGE coefficient. Notably, we find quantisation of the CPGE that persists in the presence of more than two bands, and also in the presence of non-linear corrections. We therefore demonstrate that multiplicative topological phases are an invaluable tool in pursuit of quantised topological response signatures. 

\begin{figure}
    \centering\includegraphics[width=0.9\textwidth]{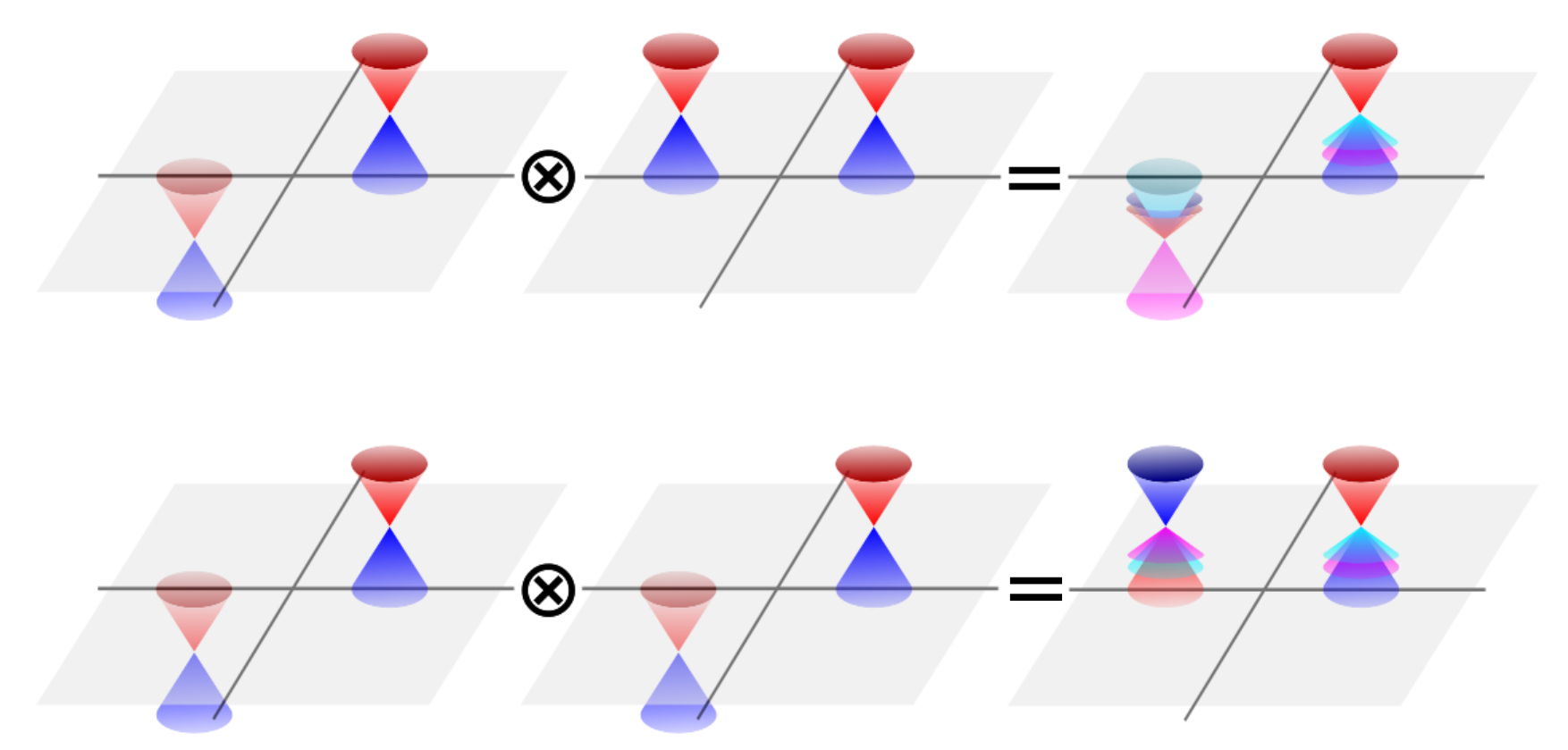}
    \caption{Schematic band structures showing the parallel multiplicative Weyl semimetal construction.
    Top: even-odd tilting, bottom: odd-odd tilting.
    All the Weyl nodes are located at $\bk = \left( 0, 0, \pm \pi/2\right)$ with the energies shifted in a $\bk$-even or odd fashion.
    The colors indicate the the different band indices.}
    \label{fig:cpgeschematic}
\end{figure}

\textit{II. Results}--

\hspace{3mm}\textit{II.1. Multiplicative Weyl semimetal and CPGE response contributions}--
The central quantity we focus on in this manuscript is the CPGE tensor $\beta$, which characterizes the injection current response to applied circularly polarized light~\cite{deJuan2017} defined through $dj_i/dt = \beta_{ij} (\omega) (\vb{E}(\omega) \times \vb{E}(-\omega))^j$.
The trace of the CPGE tensor is given in terms of the band structure by
\begin{align}
\label{eqn:cpge}
    & \operatorname{Tr} \beta(\omega) =\\ & \frac{\pi e^3}{\hbar V} \epsilon_{ijk} \sum_{\bk, n, m}
    f_{\bk, n, m} \Delta_{\bk, n, m}^i r^j_{\bk, n, m} r^k_{\bk, m, n} \delta(\hbar \omega - E_{\bk, n, m}),\nonumber
\end{align}
where $E_{\bk, n, m} = E_{\bk, n} - E_{\bk, m}$ with $E_{\bk, m}$ the band energies, $f_{\bk, n, m} = f(E_{\bk, n}) - f(E_{\bk, m})$ with $f(E)$ the Fermi function, $\Delta_{\bk, n, m}^i = (1/\hbar) \partial_i E_{\bk, n, m}$, and $r^i_{\bk, n, m} = i \bra{n, \bk} \partial_i \ket{m, \bk}$ the inter-band Berry connection.
The derivatives $\partial_i$ are with respect to the momentum component $k_i$.
After a change of variables, and using the fact that the delta function is only nonzero on closed surfaces $S_{nm}$ where $\hbar \omega = E_{\bk, n, m}$ is satisfied, we rewrite this in terms of a sum of surface integrals:
\begin{equation}
    \operatorname{Tr} \beta(\omega) = \frac{\pi e^3}{\hbar} \epsilon_{ijk} \sum_{n, m} \int_{S_{nm}} d\bk^2 \hat{n}_{\bk, n, m}^i
    f_{\bk, n, m} r^j_{\bk, n, m} r^k_{\bk, m, n},
\end{equation}
where $\hat{n}_{\bk, n, m}^i$ is the surface normal vector.
In case of a two-band system, using the identity $\vb*{\Omega}_n = i \sum_{m\neq n} \vb{r}_{nm}\times \vb{r}_{mn}$ expressing the Berry-curvature in terms of the inter-band Berry connection, and assuming that at zero temperature $f_{1,2} = 1$ for the entire surface $S_{12}$, we obtain the celebrated result $\operatorname{Tr} \beta(\omega) = \frac{i\pi e^3}{h^2} C$, showing that CPGE is quantised proportional to the Weyl-node charge~\cite{deJuan2017}.
This result remains approximately valid for twofold degenerate Weyl-nodes even when higher bands are included, provided that the gap to the other bands is large compared to $\omega$.
It is tempting to conclude that the quantization is robust for multifold Fermions as well, as long as full momentum shells of allowed transitions from all occupied bands to all unoccupied bands are available~\cite{Flicker2018, Le2020, Le2021, Ni2021}, allowing us to rewrite \eqref{eqn:cpge} in terms of Berry curvatures.
This, however, is only valid for $\vb k$-linear Hamiltonians, where the eigenstates do not depend on $|\vb{k}|$, as noted in Ref.~\onlinecite{Flicker2018}, resulting in corrections to the quantised values if higher-order terms are included in the Hamiltonian, even when no higher bands are present.

\begin{figure}[ht!]	
\begin{center}
	\includegraphics[width=1\columnwidth]{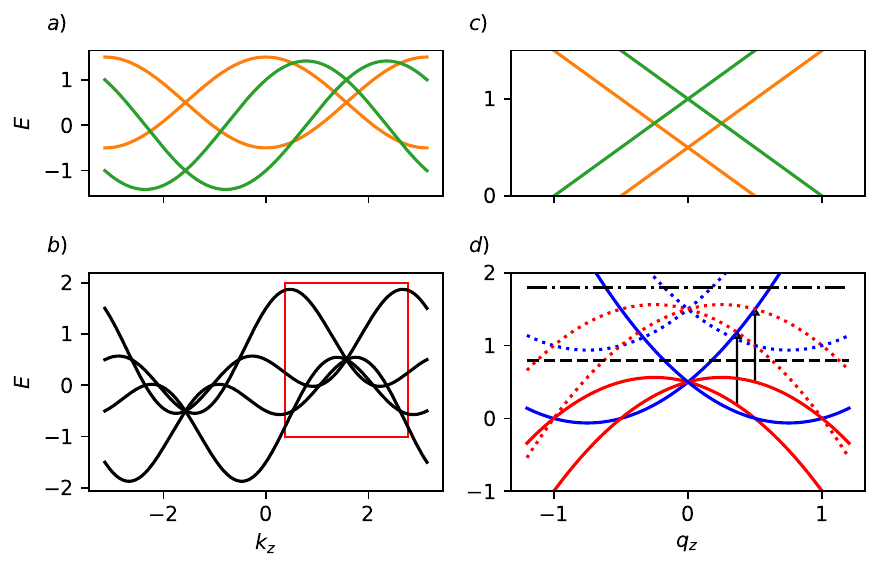}
 \end{center}
\caption{Parallel multiplicative Weyl semimetal band structures along the $k_z$ axis. Two parents $a)$ with even and odd energy shifts of the Weyl nodes (orange and green) produce the child band structure $b)$. Panels $c)$ and $d)$ show the linearized band structure around the node at $k_z = \pi/2$ (red box).Panel $d)$ shows the child band structure (solid lines where blue/red corresponds to bands with $s_1 s_2 = \pm 1$), the band structure shifted by $\omega$ (dotted), the Fermi level (dashed black) and the Fermi level shifted by $\omega$ (dash-dot black). The allowed transitions can be read off the diagram as intersections of solid and dotted lines of different color between the two black lines, marked by arrows.The parameters $t_{01} = 0.5$, $t_{02} = 1$, $\omega=1$ and $E_F=0.8$ are used in all panels.}\label{fig:band_structures_pll}
 \end{figure}

We study several variants of the MWSMs. Previously-studied examples are characterised by Bloch Hamiltonians, which are symmetry-protected tensor-products of two ``parent'' Bloch-Hamiltonians $H_1(\vb k)$ and $H_2(\vb k)$ resulting in the child $H_c(\vb k) = H_1(\vb k) \otimes H_2(\vb k)$.
We choose each parent to be a two-band Weyl-semimetal, review the general results regarding the CPGE, and in the next section we present phase-diagrams with various Weyl-node configurations.
In a MWSM with $2\times 2$ bands (two from each WSM parent), we replace the band-indices $n$ with a pair of binary indices $s_1, s_2$ each taking $\pm$ values, referring to the pair of bands in the parents.
The energies and the eigenstates factorize as $E_{\bk, s_1 s_2} = E^{(1)}_{\bk, s_1} E^{(2)}_{\bk, s_2}$, and $\ket{\bk, s_1 s_2} = \ket{\bk, s_1}_1 \otimes \ket{\bk, s_2}_2$.
As a result, the inter-band Berry connection factorizes as:
\begin{equation}
    \br_{s_1 s_2, s'_1 s'_2} = \delta_{s_1s'_1} \br^{(2)}_{s_2s'_2} +
    \delta_{s_2s'_2} \br^{(1)}_{s_1s'_1},
\end{equation}
where we suppressed the $\bk$-dependence.
Substituting this into~\eqref{eqn:cpge}, and using the fact that, with $\omega\neq 0$, terms with $n=m$ do not contribute, we find that the only remaining terms are
\begin{equation}
    r^{(1)j}_{+-} r^{(1)k}_{-+}, \;r^{(1)j}_{-+} r^{(1)k}_{+-} , \;
    r^{(2)j}_{+-} r^{(2)k}_{-+} , \; r^{(2)j}_{-+} r^{(2)k}_{+-},
\end{equation}
which can be viewed as a type of selection rule for the allowed transitions, only allowing four of the six possible transitions between the four bands, see Fig.~\ref{fig:band_structures_pll}~d).
As all of these contributions only involve the Berry connection of a single two-band parent, they are individually quantised if the transitions are allowed for entire closed momentum shells.
These contributions may cancel each other, however.
For example, if $f_{\bk, +-, --} = f_{\bk, -+, ++} = 1$ (for an entire shell of $\bk$), then the contributions from $r^{(1)j}_{+-} r^{(1)k}_{-+}$ and $r^{(1)j}_{-+} r^{(1)k}_{+-}$ cancel, and similarly for parent 2.
This is the case when the two parents are both simple Weyl-nodes at the same $\bk$ and at zero energy.

\hspace{3mm}\textit{II.2. Models and symmetry considerations}--
Our starting point is the simple two-band WSM Hamiltonian~\cite{}
\begin{equation}
    \mathcal{H}^0_{i,z} (\vb k) = (M_i(\vb{k})\sigma^z+\sin k_x\sigma^x+\sin k_y\sigma^y)
\end{equation}
where $M_i(\vb{k})=2+\gamma_i-\sum_{j=x,y,z}\cos k_j$, we reserve the index $i=1,2$ to index the parents later. 
This Hamiltonian describes a WSM with Weyl-nodes separated along the $z$ axis.
The separation is controlled by $\gamma_i$, for most of the following we choose $\gamma_i = 0$ so the Weyl-nodes are located at $\bk = \left( 0, 0, \pm \pi/2\right)$.
We analogously define $\mathcal{H}^0_{i,y}$ with Weyl-nodes separated in the $y$ direction, the only difference from $\mathcal{H}^0_{i,z}$ is that the role of $k_y$ and $k_z$ is interchanged.
Following Ref.~\onlinecite{Pal2023} (up to a change of basis), we define the MWSM parallel and perpendicular Bloch Hamiltonians:
\begin{subequations}
\label{eqn:MWSM0}
\begin{equation}
\begin{split}
\mathcal{H}^0_{\parallel}(\vb{k})&= \mathcal{H}^0_{1,z}(\vb{k})\otimes \mathcal{H}^0_{2,z}(\vb{k}),
\end{split}
\end{equation}
\begin{equation}
\begin{split}
\mathcal{H}^0_{\perp}(\vb{k})&= \mathcal{H}^0_{1,y}(\vb{k})\otimes \mathcal{H}^0_{2,z}(\vb{k}),
\end{split}
\end{equation}
\end{subequations}
with further details in the Appendix, section~\ref{sec:numerics}.

The two systems defined above have both time reversal $(\mathcal{T}^2=+1)$ and inversion symmetry.
A net non-zero quantised CPGE is only possible when inversion symmetry is broken, so that the contributions from Berry curvature around both the Weyl nodes at the same energy do not cancel \cite{deJuan2017}.
In order to break these symmetries, we introduce a \emph{tilting} term, shifting the band energies:
\begin{equation}
\begin{split}
\mathcal{H}_{i,j} &= t_{0i}f_i(k_j)\sigma^0+\mathcal{H}^0_{i,j}(\vb{k})
\end{split}
\end{equation}
where we choose $f_i$ as either an even (specifically constant $1$) or an odd (sine) function of $k_j$, and define the MWSM Hamiltonians the same way as Eqns.~\eqref{eqn:MWSM0} with the $0$ index dropped.
See Fig.~\ref{fig:band_structures_pll}~a,~b) for an example parallel MWSM band structure.

In both the parallel and the perpendicular MWSM case, if $f_1$ is odd and $f_2$ is even (or vice-versa), both inversion and time reversal symmetries are broken.
If $f_1$ and $f_2$ are both odd functions, inversion symmetry is broken but time reversal symmetry is still preserved, and as a result, the time-reversed Weyl nodes contribute the same to the CPGE, and only even quantised values are allowed.
If both parents have even tilting, inversion symmetry is maintained and the CPGE vanishes.

\begin{figure}[ht!]
	\begin{center}
		\includegraphics[width=1\columnwidth]{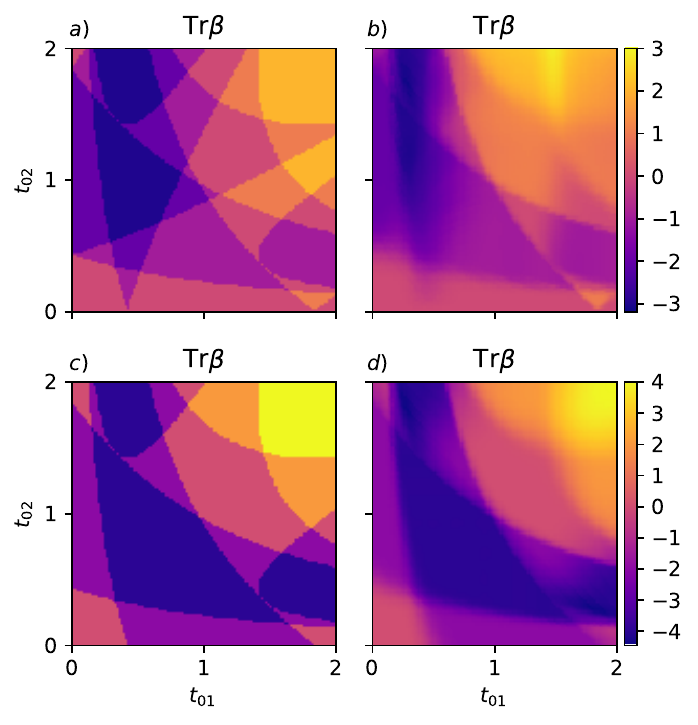}
	\end{center}
	\caption{Phase diagram of the parallel MWSM as a function of the tilting parameters of the parents. Continuum approximation (a) and numerical evaluation from the lattice model (b) in the even-odd case.
 Continuum approximation (c) and lattice model (d) in the odd-odd case.
 All phase diagrams are with parameters $\omega=1$ and $E_F=0.8$.
	\label{fig:phase_diagram_pll}}
\end{figure}

\hspace{3mm}\textit{II.4. Phase diagram}-- 
We calculate the CPGE from the lattice model by numerically evaluating eqn.~\ref{eqn:cpge}, see Appendix~\ref{sec:numerics} for details.
The resulting phase diagram is shown in Fig.~\ref{fig:phase_diagram_pll}~b).
As a function of the tilting parameters $t_{01}$ and $t_{02}$, we observe large regions with the CPGE trace quantised to high accuracy, with intervening areas of non-quantised values.
This is expected when the shift in the node energies becomes comparable to the band-width, and the allowed transitions no longer occur on closed momentum-shells as required for quantization.

As a comparison, we also calculate a semi-analytical phase diagram using a linearized continuum model.
As argued in Sec. II.1., and confirmed by the numerics on the lattice model, quadratic terms do not influence the quantization, justifying the simplification.
First we consider a single isolated Weyl point in both parents, located at the same momentum:
\begin{equation}
    H(\bk) = (\vb{q}\cdot\boldsymbol{\sigma} + t_{01} \mathbbm{1}) \otimes (\vb{q}\cdot\boldsymbol{\sigma} + t_{02} \mathbbm{1}),
\end{equation}
with $\vb q = \vb k - \vb k_0$ the distance from the Weyl point momentum, see Fig.~\ref{fig:band_structures_pll}~c,~d).
Here we set the velocity and charge of both Weyl points to $1$, with an overall energy shift determined by $t_{0i}$.
The spectrum is
\begin{equation}
    E(q)_{s_1, s_2} = (s_1 q + t_{01}) (s_2 q + t_{02}),
\end{equation}
where we use $q = |\vb{q}|$ as a result of the spherical symmetry of the problem, and $s_i=\pm 1$ specify the branch of the dispersion.
As we saw before, only transitions where one of the band indices change contribute to the CPGE.
For specificity, let us assume that $s_2$ changes (the other case can be treated by interchanging $1 \leftrightarrow 2$).
The condition for a transition is:
\begin{equation}
    E(q)_{s_1,s_2} - E(q)_{s_1,-s_2} = \omega,
\end{equation}
which can be solved for the momentum $q_0$ of the allowed transitions:
\begin{equation}
    q_0 = -s_1 \frac{t_{01}}{2} \pm \sqrt{t_{01}^2 - 2 s_1 s_2 \omega}.
\end{equation}
For this to correspond to an allowed transition from an occupied to an empty state, we also require
\begin{equation}
    E(q_0)_{s_1,s_2} < E_F < E(q_0)_{s_1,-s_2},
\end{equation}
taking the Fermi function as a step function, see Fig.~\ref{fig:band_structures_pll}~d) for an illustrative example.
The value of $\beta$ changes where one of these inequalities becomes an equality and a transition becomes allowed or forbidden.
These result in quadratic equations in $t_{01}$ and $t_{02}$, giving phase boundaries along conic sections.
The value of $\beta$ is given by summing over all of the allowed transitions (with $q_0 > 0$ to avoid double-counting), multiplying with the sign $s_1 s_2$ and the sign of the velocity difference between the two bands at $k_0$.
The resulting phase diagram, taking into account the total contribution of the two Weyl nodes treated the above way, is shown in Fig.~\ref{fig:phase_diagram_pll}~a) and c).
The two approaches result in similar phase diagrams in both the even-odd and the odd-odd tilting case, with some differences near the phase boundaries, especially for large parameters, as expected. 

\begin{figure}[t]
	\begin{center}
		\includegraphics[width=1\columnwidth]{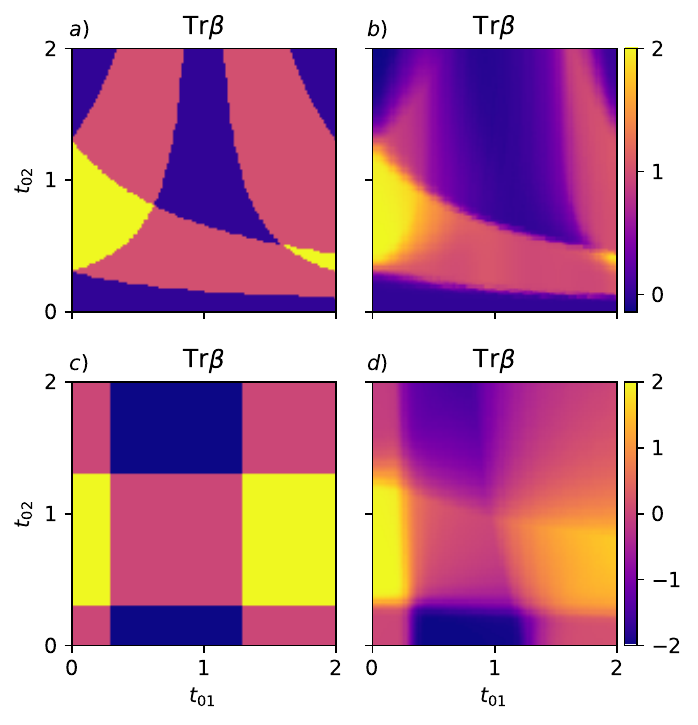}
	\end{center}
	\caption{Phase diagram of the perpendicular MWSM as a function of the tilting parameters of the parents. Continuum approximation (a) and numerical evaluation from the lattice model (b) in the even-odd case.
 Continuum approximation (c) and lattice model (d) in the odd-odd case.
 All phase diagrams are with parameters $\omega=1$ and $E_F=0.8$.
	\label{fig:phase_diagram_perp}}
\end{figure}

We also consider a perpendicular MWSM, where the role of $k_y$ and $k_z$ is reversed in the first parent relative to the second parent.
The analytical calculation of the phase diagram in this case is analogous to the parallel case.
The fact that only one parent has a Weyl-node in the vicinity of any $\bk$-point allows for simplifications. In the even-odd case, at linear order the dispersion around the Weyl-node of the second parent located at $\bk = \left( 0, 0, \pm \pi/2\right)$ is
\begin{equation}
    E(q)_{s_1, s_2} = (s_1 + t_{01}) (s_2 q \pm t_{02}),
\end{equation}
where we assumed that $\gamma_1 = \gamma_2 = 0$.
The linearized dispersion around the Weyl-nodes of the first parent located at $\bk = \left( 0, \pm \pi/2, 0\right)$ are
\begin{equation}
\label{eqn:pp_odd}
    E(q)_{s_1, s_2} = (s_1 q + t_{01}) s_2,
\end{equation}
with no energy shift proportional to $t_{02}$ at the node locations because of the odd tilting in the second parent.
We neglect the anisotropic $q_z$-linear energy shift in the second parent, because our semi-analytical treatment assumes spherical symmetry, this in essence means that we average over rotations around the Weyl-node.
The contributions to the CPGE are opposite for the two nodes, resulting from the opposite charge of the nodes.
We use the same method as for the parallel case to find the allowed transitions and their contributions for each node, the resulting total CPGE is shown in Figs.~\ref{fig:phase_diagram_perp}.
The phase diagram in the perpendicular odd-odd case is very simple, as the dispersion around all four nodes has the form of~\eqref{eqn:pp_odd}, hence the dependence of the CPGE on the tiltings completely decouples for the two parents in the low frequency (continuum) limit. 

\textit{III. Discussion and Conclusion}--In this work, we characterise the circular photogalvanic effect in multiplicative topological semimetals, or topological semimetals characterised by Bloch Hamiltonians with symmetry-protected tensor product structure. Given past work on quantisation of the CPGE for systems with Weyl nodes and multifold fermions, we explore the potential of multiplicative Weyl semimetals in realising quantised CPGE. 

For both parallel and perpendicular MWSMs with two parent Weyl semimetals as introduced in past work, we find rich phase diagrams as a function of the tilting parameters for the parents.
We obtain these results both numerically from lattice tight-binding models, and analytically using a continuum approximation, and find the same phase diagrams up to expected corrections.
The multiplicative structure yields quantisation in systems with more than two bands even in the presence of non-linear momentum dependence over large regions of phase space, revealing an approach to more robust quantisation than realized previously.
This demonstrates the potential of multiplicative topological states in realizing quantised observables and exotic optical signatures promising for applications.
Furthermore, our work demonstrates that CPGE is a versatile probe in characterising  complex topology, distinguishing topological degeneracies of multiplicative topological semimetals from ostensibly similar topological degeneracies according to the degree of the degeneracy and net topological charge, such as Dirac nodes.

\begin{acknowledgments}
We thank J. Winter for helpful discussions. This research was supported in part by the National Science Foundation under Grants No.NSF PHY-1748958 and PHY-2309135, and undertaken in part at Aspen Center for Physics, which is supported by National Science Foundation grant PHY-2210452.
D.V. was supported by the Deutsche Forschungsgemeinschaft (DFG, German Research Foundation) under Germany’s Excellence Strategy through the W\"urzburg-Dresden Cluster of Excellence on Complexity and Topology in Quantum Matter – ct.qmat (EXC 2147, project-id 57002544).
\end{acknowledgments}

\bibliography{bibliography.bib}
\clearpage
\appendix

\onecolumngrid
\section{Lattice Hamiltonians}
\label{sec:lattice_hams}
For the MWSM parallel case,
\begin{equation}
\begin{split}
H_\parallel(\boldsymbol{k})=&\left[t_{01}\tau^0+\left(2+\gamma_1-\sum_{i} \cos k_i\right)\tau^z +\sin k_x\tau^x+\sin k_y\tau^y\right] \\
&\otimes \left[t_{02}\sin k_z\sigma^0-\left(2+\gamma_2-\sum_{i} \cos k_i\right)\sigma^z -\sin k_x\sigma^x+\sin k_y\sigma^y\right].
\end{split}
\end{equation}
For the MWSM perpendicular case with inversion along $k_z$,
\begin{equation}
\begin{split}
H_\perp(\boldsymbol{k})=&\left[t_{01}\tau^0+\left(2+\gamma_1-\sum_{i} \cos k_i\right)\tau^z+\sin k_x\tau^x+\sin k_z\tau^y\right]\\
&\otimes \left[t_{02}\sin k_z\sigma^0-\left(2+\gamma_2-\sum_{i} \cos k_i\right)\sigma^z-\sin k_x\sigma^x+\sin k_y\sigma^y\right].
\end{split}
\end{equation}
\twocolumngrid
\section{Numerical evaluation of CPGE}
\label{sec:numerics}
We evaluate eqn. \eqref{eqn:cpge} using finite difference derivatives with $\delta k_d = 10^{-9}$, and gauge-fixed eigenstates that maximize the real part of the overlap between eigenstates at neighboring $k$-points.
We use a cubic grid with spacing $\delta k_i = 2\pi/N_k$ with $N_k = 70$ for the integral.
The delta-function is replaced with a finite-width Gaussian of width $\delta k_g = 0.1$.

\end{document}